\documentclass[%
reprint,
showkeys,
nofootinbib,
amsmath,amssymb,
aps,
prl,
]{revtex4-1}
\usepackage[marginal]{footmisc}
\usepackage{graphicx}
\usepackage{dcolumn}
\usepackage{bm}
\usepackage[colorlinks,allcolors=blue,CJKbookmarks=True]{hyperref}

\begin{document}

\title{
Q-balls from thermal balls during a first-order phase transition: a numerical study}

\author{Yuan-Jie Li}
\email{liyuanjie23@mails.ucas.ac.cn}
\affiliation{%
International Centre for Theoretical Physics Asia-Pacific, University of Chinese Academy
of Sciences, 100190 Beijing, China
}%
\affiliation{%
Taiji Laboratory for Gravitational Wave Universe, University of Chinese Academy of Sciences, 100049 Beijing, China
}%
\author{Jing Liu}
\email{liujing@ucas.ac.cn}
\affiliation{%
International Centre for Theoretical Physics Asia-Pacific, University of Chinese Academy
of Sciences, 100190 Beijing, China
}%
\affiliation{%
Taiji Laboratory for Gravitational Wave Universe, University of Chinese Academy of Sciences, 100049 Beijing, China
}%
\author{Zong-Kuan Guo}
\email{guozk@itp.ac.cn}
\affiliation{%
Institute of Theoretical Physics, Chinese
Academy of Sciences, P.O. Box 2735, Beijing 100190, China
}%
\affiliation{%
School of Physical Sciences, University of Chinese Academy of Sciences, No.19A Yuquan
Road, Beijing 100049, China
}%
\affiliation{%
School of Fundamental Physics and Mathematical Sciences, Hangzhou Institute for Advanced Study, University of Chinese Academy of Sciences, Hangzhou 310024, China
}%

\begin{abstract}
We numerically study the Q-ball formation triggered by a cosmological first-order phase transition within the Friedberg-Lee-Sirlin model. By performing lattice simulations, we track the nonequilibrium dynamics throughout the transition, providing a precise description of the Q-ball formation mechanism and the resulting mass spectrum.
Collapsing false-vacuum regions first form thermal balls, which subsequently cool via dissipative interactions and stabilize into long-lived Q-balls with nonzero spin. 
We observe a large population of low-mass Q-balls, as well as rare, massive Q-balls that are several times larger than the analytical prediction.
The final Q-ball population exhibits a broad mass spectrum spanning over two orders of magnitude, characterized by an exponential tail of number density at large masses. 
The simulations suggest that the Q-ball abundance is approximately $50\%$ higher than predicted by analytical estimates, adjusting the result in the context of Q-balls as dark matter candidates.
\end{abstract}

\maketitle

\emph{Introduction.} Cosmological first-order phase transitions~(FOPTs) play a crucial role in the evolution of the early Universe, marking epochs of symmetry breaking as the vacuum structure of fundamental fields changed during cosmic cooling~\cite{Kolb:1990vq,Mazumdar:2018dfl}. FOPTs can impact the evolution of the universe through multiple channels, such as generating the baryon asymmetry~\cite{Sakharov:1967dj,Cohen:1993nk,Morrissey:2012db,Liu:2024mdo,Di:2024gsl}, seeding primordial magnetic fields~\cite{Vachaspati:1991nm,Grasso:2000wj,Widrow:2002ud,Di:2020kbw,Yang:2021uid,Di:2025ncl,Liu:2025xxb}, and leaving behind relics ranging from topological defects (domain walls, cosmic strings, monopoles)~\cite{Kibble:1976sj, Vilenkin:2000jqa,Wu:2022tpe,Wu:2022stu,Li:2023yzq,Li:2023gil,Li:2025gld,Zhao:2025kvm} to compact objects such as primordial black holes (PBHs) from large density inhomogeneities~\cite{Baker:2021nyl,Liu:2021svg}.  Particularly, strongly FOPTs proceed via bubble nucleation and can source stochastic gravitational-wave (GW) backgrounds, a key science target for GW observatories such as LISA/DECIGO/BBO and an important probe of beyond-Standard-Model physics~\cite{Hogan:1986qda,Kosowsky:1992rz,Kamionkowski:1993fg,Crowder:2005nr,Kawamura:2006up,Caprini:2015zlo,Hindmarsh:2015qta,AmaroSeoane:2017pwj,Caprini:2019egz,Wang:2025eee,Zou:2025sow}.

After an FOPT, the true vacuum typically exhibits a spontaneously broken symmetry, and other fields acquire masses via the well-known Higgs mechanism.  This mass-gaining mechanism can also enable the formation of long-lived solitonic relics~\cite{Kawana:2022lba,Bai:2022kxq,Xie:2024mxr,Zhou:2024mea,Kanemura:2025ixp}. In particular, the Friedberg-Lee-Sirlin (FLS) model provides a minimal and well-controlled setting in which a conserved global $U(1)$ charge and an additional scalar field allow energetically favored, localized charge lumps to exist, known as Q-balls~\cite{Friedberg:1976me, Coleman:1985ki, Loiko:2019ydl}. Q-balls have been studied in early-Universe as dark matter candidates, potential GW sources, and progenitors of PBHs~\cite{Kusenko:1997si, Carr:2020gox, Dent:2025bqk,Chen:2025oxo}, where most discussions concentrate on the Affleck-Dine mechanism of Q-ball generation~\cite{Affleck:1984fy,Enqvist:1999mv,Kasuya:2000wx,Kasuya:2001hg,Tsumagari:2009na,Multamaki:2002hv,Enqvist:2003gh,Zhou:2015yfa,White:2021mfb,Hou:2022jcd}.

While the properties of FLS Q-balls can be qualitatively understood analytically~\cite{Friedberg:1976me,Krylov:2013qe,Loiko:2019ydl,Jiang:2024zrb}, their nonlinear production process during an FOPT lacks dedicated studies.  The highly out-of-equilibrium dynamics of an FOPT also prevents reliable analytic predictions for their production efficiency and mass spectra.  In this work, we therefore perform numerical simulations of a FOPT in the FLS model and explicitly demonstrate the emergence of Q-balls from the FOPT dynamics, providing the first numerical realization of this production channel.  We find an abundance of low-mass Q-balls arising from the fragmentation of the false vacuum, and also observe spinning Q-balls carrying nonzero angular momentum.  We find that the resulting mass function is broad, and the total abundance obtained from simulations can adjust the dark matter density in scenarios where Q-balls constitute dark matter. Furthermore, under suitable conditions, sufficiently heavy Q-balls are expected to collapse into PBHs, allowing us to derive a corresponding PBH mass spectrum. This framework also enables more precise predictions for the detection or constraints on Q-balls as compact dark matter candidates.

\emph{Model.} One of the simplest frameworks admitting Q-balls is the FLS model \cite{Friedberg:1976me}, which consists of a complex scalar field $\chi$ and a real scalar field~$\phi$,
\begin{align}
\mathcal{L}
=
\frac{1}{2}(\partial_\mu \phi)^2
+&
(\partial_\mu \chi)^{\ast}(\partial^\mu \chi)
-
U(\phi,\chi),\\
U(\phi,\chi)
=&
U(\phi)
+
g^2 \phi^2 \chi^{\ast}\chi,
\end{align}
where $U(\phi)$ denotes the effective potential of $\phi$.

The field $\chi$ carries a conserved global $U(1)$ charge associated with the transformation
$\chi \to e^{i\alpha}\chi$,
and its effective mass is generated through the interaction with $\phi$.
As a result, $\chi$-particles are massless in regions where $\phi\simeq0$, whereas they become massive in the true vacuum where $\phi$ acquires a nonzero expectation value $m_\chi^2(\phi)=g^2\phi^2$.
This mass hierarchy provides a natural trapping mechanism: $\chi$-particles are energetically disfavored to leave regions of small $\phi$, leading to the formation of localized bound states of false vacuum.

The conserved $U(1)$ symmetry implies a conserved Noether charge
\begin{equation}
Q
=
\int d^3x\, i\left(\chi^\ast \pi_\chi^\ast - \chi \pi_\chi \right)
=\int\mathrm{d}^3x\,2\,\mathrm{Im}[\chi\pi_{\chi}],
\end{equation}
where $\pi_\chi \equiv \partial \mathcal{L}/\partial \dot{\chi}$ denotes the canonical momentum conjugate to $\chi$. Q-balls correspond to field configurations that minimize the total energy at fixed $Q$.
Such configurations are stable against decay into free $\chi$-particles provided that their energy satisfies $E(Q)<m_\chi^{\rm vac}Q$, where $m_\chi^{\rm vac}$ denotes the $\chi$ mass in the true vacuum.

We take the effective potential of the $\phi$ field to be
\begin{align}
U(\phi)
=
\frac{1}{4}\lambda \phi^4
-
\frac{\beta}{3}\phi^3
+
\frac{1}{2}\mu_\phi^2 \phi^2+U_0,
\end{align}
where $U_0$ corresponds to the false-vacuum energy density. The interaction term $g^2 \phi^2 \chi^{\ast}\chi$ does not change the vacuum expectation value for the $\chi$ field, therefore, the false and true vacua are located at
$(\phi,\chi)=(0,0)$ and $(\phi_{\mathrm{b}},0)$, respectively, where $\phi_{\mathrm{b}}$ denotes the broken phase of the $\phi$ field.

The reduced effective potential of $\phi$, which governs the tunneling rate of true vacuum bubbles~\cite{Li:2025zxa}, is lifted by $\chi$-field fluctuations
\begin{align}
U_{\mathrm{re}}(\phi)
=
\frac{1}{4}\lambda \phi^4
-
\frac{\beta}{3}\phi^3
+
\frac{1}{2}M_\phi^2 \phi^2+U_0 ,
\end{align}
where the effective mass parameter is given by
$M_\phi^2=\mu_\phi^2+2g^2\langle|\delta\chi|^2\rangle$ applying the mean-field approximation, and $\langle\cdots\rangle$ denotes a spatial average. The field value at the minimum of $U_{\mathrm{re}}(\phi)$ is denoted by $\eta$.

We also account for the friction of background plasma on bubble walls, 
and the resulting equation of motion can be written in covariant form as
\begin{equation}
-\Box \phi-\frac{\partial U(\phi,\chi)}{\partial\phi}
=
\Gamma\,\gamma\left(\dot{\phi}+V^i\partial_i\phi\right),
\label{eq:eom_fluid}
\end{equation}
where $V^i$ is the three-velocity of the fluid, $\gamma=(1-\mathbf{V}^2)^{-1/2}$, and $\Gamma$ parametrizes the strength of dissipation.
In this work we focus on the scalar-field dynamics and neglect the bulk motion of the fluid, setting $V^i=0$ (hence $\gamma=1$)~\cite{Moore:1995ua,John:2000zq,Cutting:2018tjt}.

\emph{Numerical setup}. 
We solve the equations of motion on a three-dimensional cubic lattice with periodic boundary conditions using the open-source package \textit{\texttt{CosmoLattice}} (v1.2~\cite{Figueroa:2020rrl,Figueroa:2021yhd}). Spatial derivatives are evaluated with a second-order central finite-difference scheme, and time evolution is performed using the leapfrog algorithm. The simulations employ \(N=512\) grid points per spatial dimension. The lattice spacing \(\Delta x\) is chosen such that the box size is \(L \equiv N\Delta x = 128\,\omega_*^{-1}\), where \(\omega_* \equiv \sqrt{\lambda}\,\eta\) sets the characteristic mass scale. The time step is fixed to \(\Delta t = 0.2\,\Delta x\) to ensure numerical stability. Throughout this work, the dissipation coefficient in Eq.~\eqref{eq:eom_fluid} is set to \(\Gamma = 0.1\,\omega_*\).

The critical bubble profile is obtained using \textit{\texttt{FindBounce}}, which solves the bounce equation in the reduced potential \(U_{\rm re}(\phi)\) via the shooting method~\cite{Guada:2020xnz}.
To trigger the FOPT, we nucleate \(N_\mathrm{b}\) critical bubbles at the initial time, requiring the centers of any two bubbles to be separated by more than \(2R_{\mathrm{c}}\) in order to avoid overlap, where $R_\mathrm{c}$ denotes the radius of the critical bubble. To initialize multiple bubbles without introducing sharp cutoffs, we superimpose individual critical-bubble profiles in a smooth manner (see Appendix for details).

In the simulations, we fix the model parameters to \(\beta = 1.3\,\omega_*^2\eta^{-1}\), \(M_\phi^2 = 0.3\,\omega_*^2\), and \(g^2 = 10\lambda\). For this choice, the critical bubble has a radius \(R_{\mathrm{c}} = 10.06\,\omega_*^{-1}\) and a wall thickness \(l_0 = 3.05\,\omega_*^{-1}\). 

The complex scalar field \(\chi\) is decomposed into two real components, \(\chi = (\chi_1 + i\chi_2)/\sqrt{2}\). The homogeneous modes are set to zero, and the inhomogeneous fluctuations are initialized as Gaussian random fields corresponding to vacuum fluctuations. A net \(U(1)\) charge is introduced by appropriately initializing the canonical momentum conjugate to \(\chi\). Further details of the initialization procedure, including the treatment of vacuum fluctuations and the injection of a nonzero \(U(1)\) charge, are provided in the Appendix.

\emph{Results.} We present the results of our numerical simulations, including a detailed evolution of single Q-ball, and the predictions for the Q-ball mass spectrum obtained from an ensemble of realizations of FOPTs.
For a single Q-ball, an spherical false-vacuum region is initialized at the center of the simulation box with $U(1)$ charge injected into $\chi$-field, see the Appendix for details.
The size of the central false-vacuum region is characterized by an effective radius defined through its volume,
$R(t)=\left(3V_{\rm ball}(t)/4\pi\right)^{1/3}$.

\begin{figure}[htb]
    \centering
    \includegraphics[width=\linewidth]{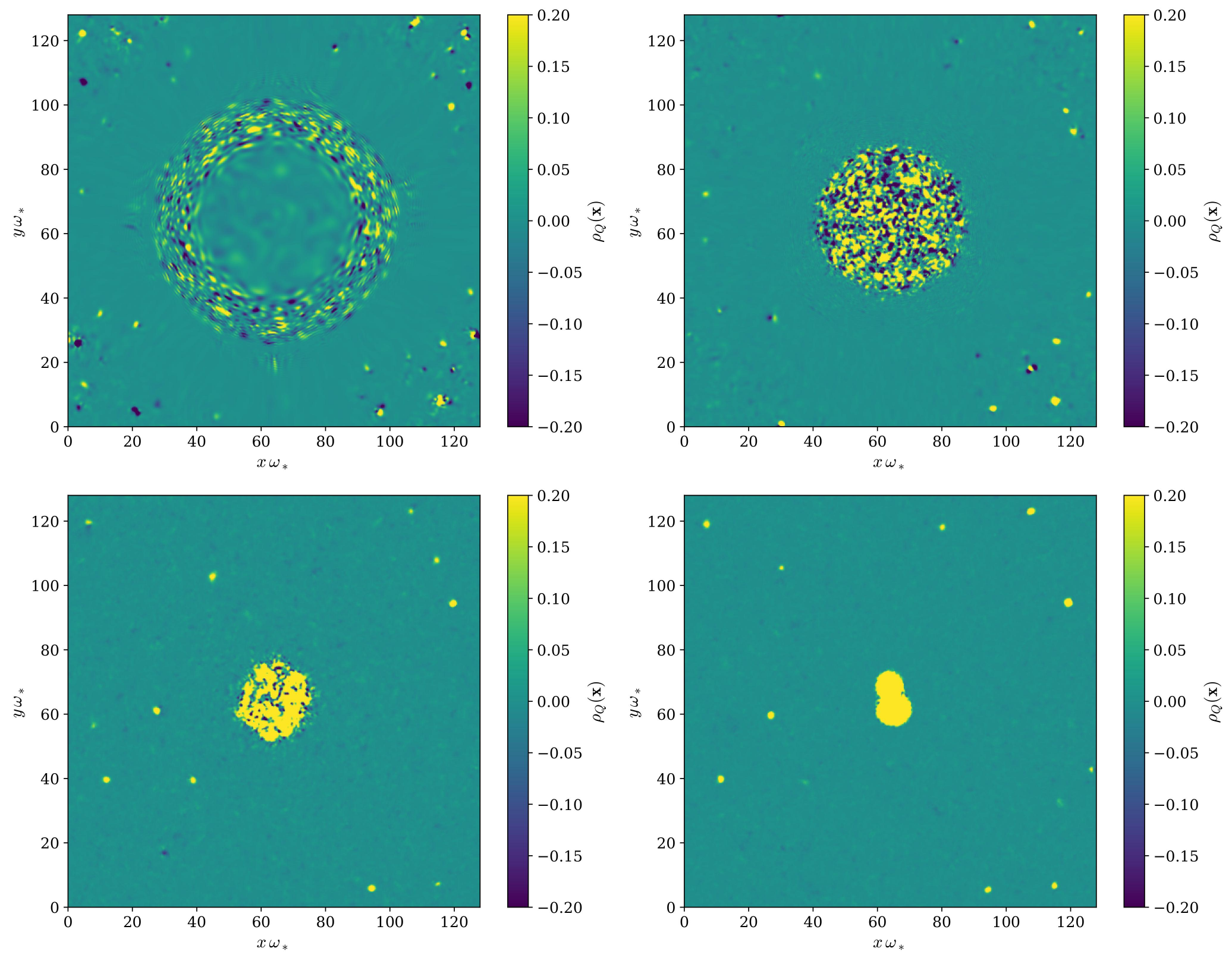}
    \caption{Formation, cooling, and stabilization of a single Q-ball from a collapsing false-vacuum bubble.}
    \label{fig:single_ball}
\end{figure}

Figure~\ref{fig:single_ball} illustrates the contraction of an initially spherical false-vacuum region and its relaxation into a Q-ball, showing slices of the $U(1)$ charge density at
$t=40\,\omega_*^{-1}$, $120\,\omega_*^{-1}$, $500\,\omega_*^{-1}$, and $1000\,\omega_*^{-1}$.
At early times, the false vacuum undergoes rapid contraction, during which $\chi$ particles are driven toward the center by the bubble wall.
As particles accumulate in the central region, their kinetic energy provides an effective pressure that temporarily halts further contraction, forming an intermediate thermal ball. Notably, at this stage, the thermal ball contains a mixed distribution of positive and negative charges in $\chi$, which marks a clear distinction from a Q-ball.
Subsequently, the friction term in Eq.~\eqref{eq:eom_fluid} leads to the dissipation of the kinetic energy of $\chi$ and the annihilation of opposite Q-charges, thus gradually cooling the thermal ball. The little pots outside the thermal ball, illustrated in Fig.~\ref{fig:single_ball}, indicate that a small portion of $\chi$ particles can pass through the bubble walls and subsequently condense into low-mass Q-balls.

At late times, the kinetic energy is largely dissipated, leaving a positively charged Q-ball whose stability is ensured by the conservation of the global $U(1)$ charge.
Its stable, non-spherical, dumbbell-like shape is consistent with a spinning Q-ball configuration~\cite{Volkov:2002aj,Kleihaus:2005me,Loiko:2018mhb}.
The contraction of the false vacuum bubble compresses the angular momentum of nearly the entire space into the Q-ball, endowing it with significant spin.

\begin{figure}[htb]
    \centering
    \includegraphics[width=0.75\linewidth]{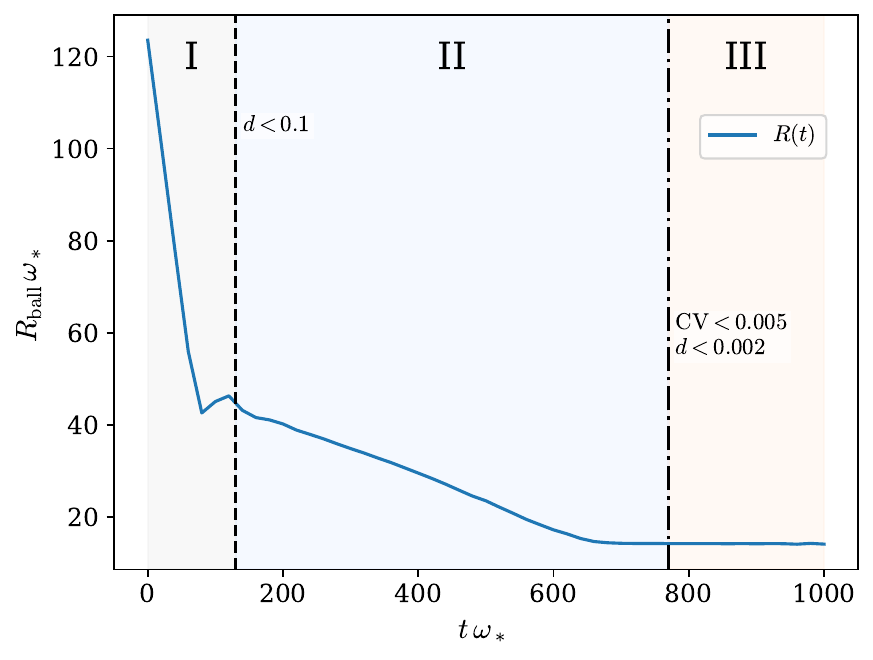}
    \caption{Time evolution of the effective radius of the central false-vacuum region. The rapid initial decrease corresponds to the contraction of the false vacuum and the formation of a thermal ball, followed by a cooling phase and the eventual emergence of a stable Q-ball.}
    \label{fig:evolution_center_bubble_radius}
\end{figure}

Figure~\ref{fig:evolution_center_bubble_radius} shows the corresponding time evolution of $R(t)$.
An initial phase of rapid contraction signals the formation of a thermal ball, followed by a cooling phase in which the radius decreases approximately linearly, consistant with the theoretical prediction~\cite{Kawana:2022lba}.
For $t\gtrsim770\,\omega_*^{-1}$, the evolution totally stabilizes, indicating the completion of cooling and the emergence of a stable Q-ball. 

We then turn to the realistic scenario of Q-ball formation during an FOPT involving multiple bubbles.
We initialize $N_{\mathrm{b}}=16$ true-vacuum bubbles with randomly distributed centers and inject a net $U(1)$ charge.
We calculate the equations of motion study Q-ball production induced by bubble expansion, collisions, and post-transition field dynamics.

\begin{figure}[htb]
    \centering
    \includegraphics[width=0.6\linewidth]{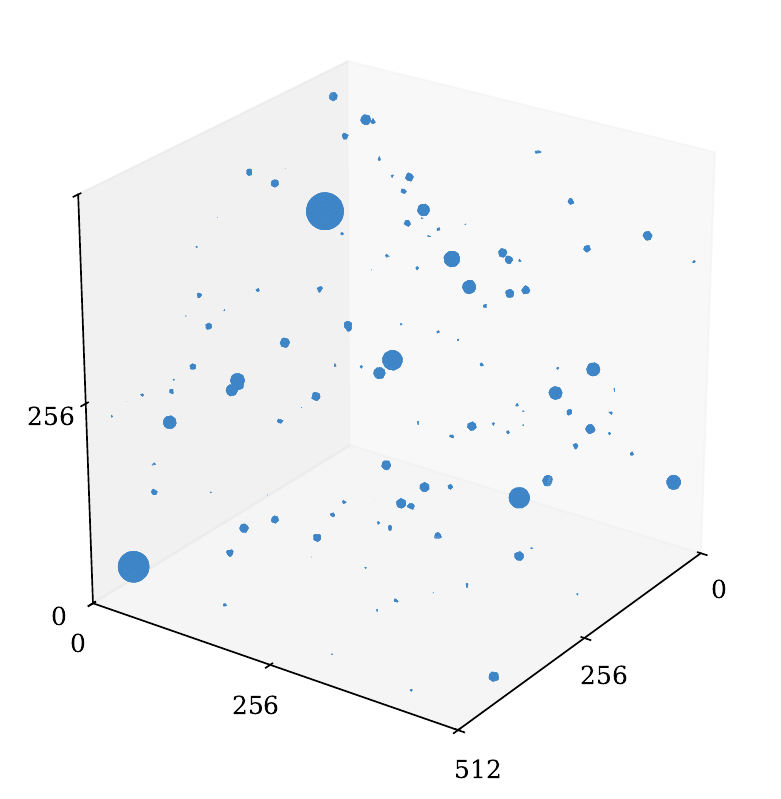}
    \caption{Isosurface visualization of Q-balls at the final simulation time $t=1000\,\omega_*^{-1}$ in a representative realization.
Blue regions indicate domains with $\phi<\phi_{\mathrm{b}}/2$, corresponding to the interiors of individual Q-balls.}
    \label{fig:isosurface_1000}
\end{figure}

We perform 30 simulations with independent random initial conditions and identify a total of 3,352 Q-balls at the final simulation time $t=1000\,\omega_*^{-1}$.
Figure~\ref{fig:isosurface_1000} shows a representative snapshot of the Q-balls formed in a single realization at the final time.
The blue regions indicate domains where $\phi<\phi_{\mathrm{b}}/2$, corresponding to the interiors of Q-balls.
We find that the majority of Q-balls exhibit no significant angular momentum and are therefore approximately spherical in shape.

\begin{figure}[htb]
    \centering
    \includegraphics[width=0.75\linewidth]{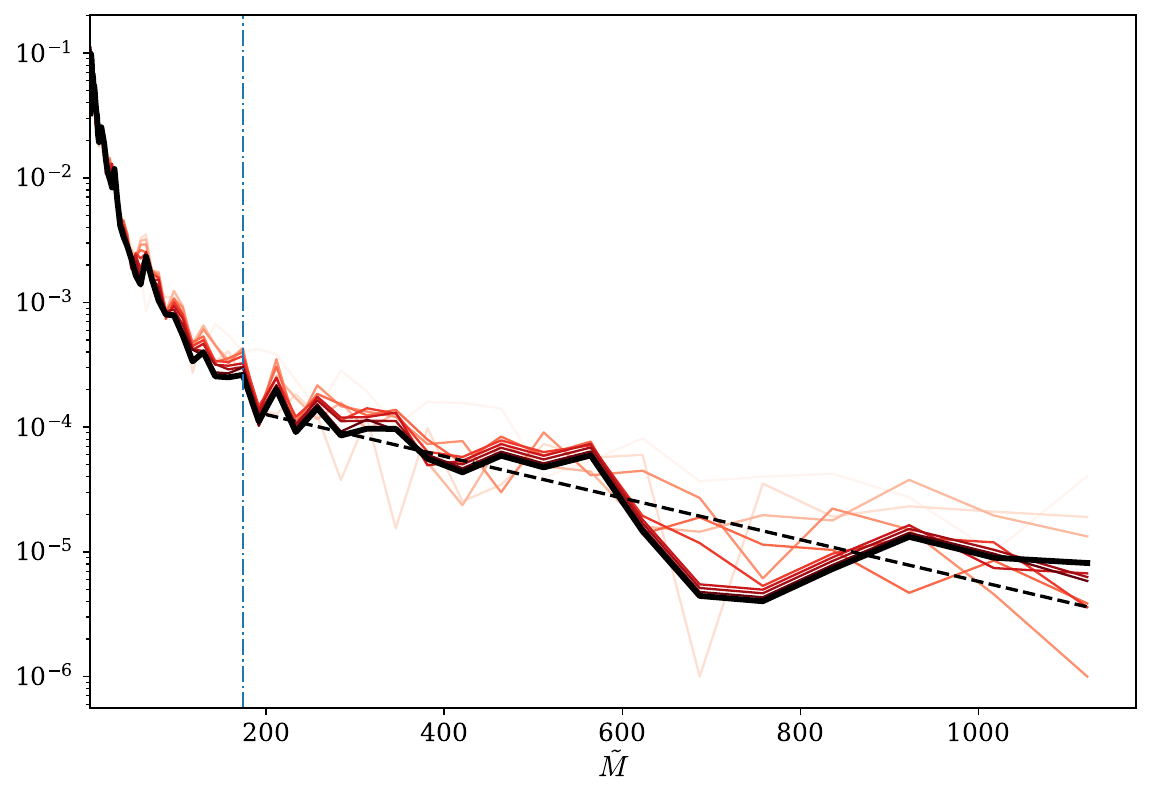}
    \caption{Histogram of the Q-ball mass.
Curves with colors ranging from light to dark correspond to time slices from $t=100\,\omega_*^{-1}$ to $t=1000\,\omega_*^{-1}$ with a time interval $\delta t=100\,\omega_*^{-1}$.
The black dashed curve shows the exponential fit given in Eq.~(\ref{eq:exp}).A light-blue dash-dotted line marks the analytically predicted characteristic Q-ball mass,$\tilde{M}_*=173.8$, in our benchmark model.}
    \label{fig:M_evolution}
\end{figure}

Figure~\ref{fig:M_evolution} shows the time evolution of the Q-ball mass distribution, where large false-vacuum regions progressively fragment into smaller domains, leading to the production of increasingly lighter Q-balls.

At late times, the mass distribution approaches a stable form with an exponential tail,
\begin{equation}
P(\tilde{M})=\exp\!\left(A-B\,\tilde{M}\right),\qquad 
\tilde{M}\equiv\frac{M\omega_*}{\eta^2},
\label{eq:exp}
\end{equation}
where $P(\tilde{M})\,\mathrm{d}\tilde{M}$ denotes the number fraction of Q-balls in the interval $(\tilde{M},\tilde{M}+\mathrm{d}\tilde{M})$.
For our benchmark model, fitting the spectrum for $\tilde{M}>200$ yields
$A=-8.21$ and $B=3.85\times10^{-3}$. This exponential behavior originates from the exponential distribution of false-vacuum volumes at the percolation stage of the FOPT.
The extended mass spectrum allows for the rare formation of supermassive Q-balls.
The blue dot-dashed line in Fig.~\ref{fig:M_evolution} shows the Q-ball mass predicted by the analytical model, $\tilde{M}_*=173.8$ for our parameters, which implies the Q-ball number density can be described by an exponential function for Q-balls larger than the analytical prediction. 
Combined with the the Q-ball mass-to-radius ratio $M_{\rm ball}/R_{\rm ball}\propto Q^{1/2}\propto M_{\rm ball}^{2/3}$~\cite{Friedberg:1976me}, the simulation results also suggests that sufficiently massive Q-balls may  collapse into PBHs without introducing other interactions.

At the low-mass end, Q-balls are produced with a significantly higher abundance, and the mass distribution deviates from a simple power-law behavior.
This reflects the preferential fragmentation of large false-vacuum regions into many smaller domains during the FOPT, as well as an additional contribution from $\chi$ particles that pass through the expanding bubble walls and subsequently condense into low-mass Q-balls, as illustrated in Fig.~\ref{fig:single_ball}.

\begin{figure}[htb]
    \centering
    \includegraphics[width=0.75\linewidth]{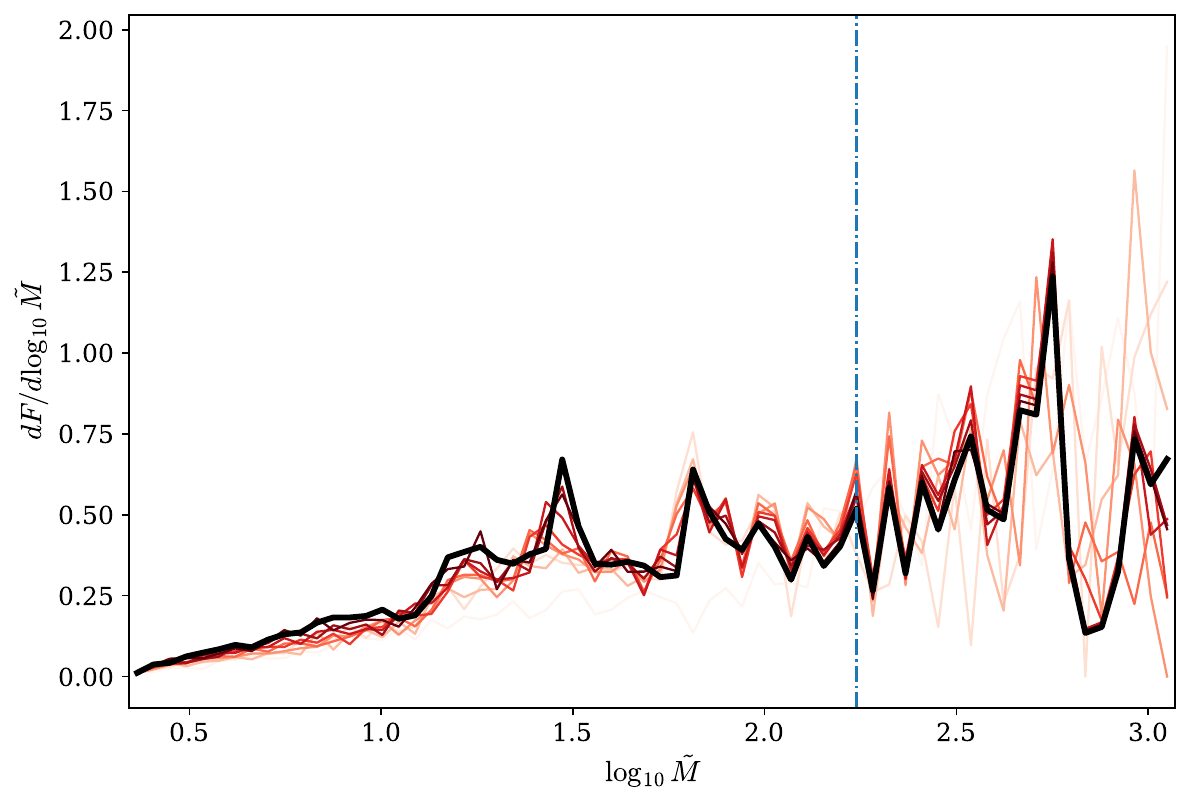}
    \caption{Time evolution of the Q-ball mass-fraction distribution $Mp(M)$.
Curves with colors ranging from light to dark correspond to time slices from early to late times.
The blue dash--dotted line marks the analytically predicted characteristic Q-ball mass.}
    \label{fig:massfrac_evolution}
\end{figure}

Figure~\ref{fig:massfrac_evolution} shows the time evolution of the mass-fraction distribution $dF/d\log_{10}M$, defined as the fractional abundance of Q-balls per logarithmic mass interval, with the total Q-ball abundance $F$ normalized to unity. Q-balls with masses below the analytically predicted characteristic value contribute $54.1\%$ of the total mass in our simulations, which implies the analytically predicted characteristic mass provides a suitable reference scale for the average behavior. However, the mass spectrum is notably broad, spanning approximately two orders of magnitude.  While low-mass Q-balls are numerous, the high-mass tail also contributes non-negligibly to the total mass budget. The high-mass end exhibits some uncertainty due to low statistical frequency. Although our simulation may not capture some extremely rare and large Q-balls, all of the Q-charge in the entire volume can at most render mass that exceeds the rightmost value on the horizontal axis in Figure~\ref{fig:massfrac_evolution} by about 0.5 in $\log_{10}M$. Thus, our simulation still covers the vast majority of the possible Q-ball mass range. To the best of our knowledge, this work provides the first numerical result of the Q-ball mass distribution.

To assess the cosmological implications of Q-balls as a dark matter candidate, we compute their abundance based on our simulations and compare it with standard analytic estimates derived from a percolation-based argument.

Approximating the Q-ball number density by the density of isolated false-vacuum regions at percolation, $n_{\rm ball}\simeq N_b/V$, and assuming efficient charge collection, the analytic prediction for the final Q-ball mass density reads~\cite{Krylov:2013qe}
\begin{equation}
\rho_{\rm ball}^{\rm (ana)} \equiv M_*\,n_{\rm ball}
\simeq \frac{4\sqrt{2\pi}}{3}\,\rho_{U(1)}^{\,3/4}
\left(\frac{N_\mathrm{b}\,U_0}{V}\right)^{1/4},
\label{eq:rho_ana_main}
\end{equation}
where $M_*$ is the typical mass of Q-ball, $\rho_{U(1)}$ is the spatially averaged \(U(1)\) charge density.
For our benchmark parameters, this yields $\rho_{\rm ball}^{\rm (ana)}
\simeq 1.3\times10^{-3}\,\omega_*^2\eta^2$.

From the simulations, we directly measure the late-time Q-ball density, and compare with $\rho_{\rm ball}^{\rm (ana)}$
\begin{equation}
r \equiv \frac{\rho_{\rm ball}^{\rm (num)}}{\rho_{\rm ball}^{\rm (ana)}} = 1.50\pm0.24 \,.
\end{equation}
The numerical result is in good agreement with the analytical prediction of the total dark matter abundance. The enhancement may originate from the effect of the extended Q-ball mass spectrum, which are not captured by the analytic monochromatic mass distribution.

Including the correction factor $r\sim1.5$ inferred from our simulations, the observed dark matter abundance can be reproduced for representative FOPT parameters without fine tuning.
For instance, an FOPT at the electroweak scale with $T_c\simeq 100~\mathrm{GeV}$, a percolation volume $V_*^{1/3}\sim4.8\times10^5\,T_c^{-1}$, and a vacuum energy scale $U_0^{1/4}\simeq 100~\mathrm{GeV}$ requires an $U(1)$ charge asymmetry $\Delta_\chi\simeq 7\times10^{-11}$ to provide observed dark matter abundance, where $\Delta_\chi$ is comparable to the baryon asymmetry of the Universe.
Alternatively, a lower-scale FOPT with $T_c\simeq 10~\mathrm{GeV}$, $V_*^{1/3}\simeq 5.2\,T_c^{-1}$, and $U_0^{1/4}\sim 10~\mathrm{GeV}$ yields the correct relic abundance for a much smaller asymmetry $\Delta_\chi\simeq 1.5\times10^{-14}$.


\emph{Conclusion.} We have presented the first numerical demonstration that Q-balls can form dynamically during a cosmological FOPT.
In the FLS model, collapsing false-vacuum regions first form thermal balls, which cool through dissipation into stable Q-balls protected by global $U(1)$ charge conservation.
The resulting population of Q-balls exhibits a rather broad mass spectrum spanning over two orders of magnitude, which is a significant departure from the monochromatic profile predicted by analytic estimates. In particular,  We observe a large number of low-mass Q-balls, alongside a small number of Q-balls that are substantially heavier than analytically expected.  
An overabundance of low-mass Q-balls arises from false-vacuum fragmentation and particle condensation during the FOPT, while the extremely large Q-balls may subsequently collapse into PBHs. Our precise numerical results yield a calculable correction to the predicted Q-ball abundance as a dark matter candidate, giving a value approximately $50\%$ higher than that obtained from analytical estimates
\section*{Acknowledgments}
This work is supported in part by the National Natural Science Foundation of China under Grants No. 12235019
and No. 12475067, in part by the National Natural Science Foundation of China under Grants No. 12147103, No. 12235019 and No.12075297.

\bibliography{ref}
\section*{Appendix}

\subsection*{I. Initialization and numerical setup}
In this Appendix, we provide additional details of the numerical implementation used in the simulations, including the initialization of vacuum fluctuations and the procedure employed to inject a net \(U(1)\) charge.

\subsection*{I.A Initialization of critical bubbles}
To initialize multiple bubbles without introducing sharp cutoffs, we superimpose the individual critical-bubble profiles as
\begin{equation}
\phi(\mathbf{x})=
\left[\sum_{i=1}^{N_b}\phi_i^2\!\left(|\mathbf{x}-\mathbf{x}_i|\right)\right]^{1/2},
\end{equation}
where \(\phi_i(r)\) denotes the spherically symmetric critical-bubble profile centered at \(\mathbf{x}_i\).

\subsection*{I.B Vacuum fluctuations of the complex scalar field}

The complex scalar field \(\chi\) is decomposed into two real components,
\begin{equation}
\chi(\mathbf{x})=\frac{1}{\sqrt{2}}\bigl(\chi_1(\mathbf{x})+i\,\chi_2(\mathbf{x})\bigr).
\end{equation}
The homogeneous modes are set to zero,
\(\langle \chi_1\rangle=\langle \chi_2\rangle=0\),
and the inhomogeneous fluctuations of each component are initialized as Gaussian random fields with a vacuum power spectrum corresponding to a massless scalar field,
\begin{equation}
\mathcal{P}_{\chi_1}(k)=\mathcal{P}_{\chi_2}(k)=\frac{1}{k},
\end{equation}
up to an ultraviolet cutoff \(k_{\max}\).

With this choice, the variance of each real component is given by
\begin{equation}
\langle \chi_1^2\rangle=\langle \chi_2^2\rangle
=\int d\ln k\,\frac{k^3}{2\pi^2}\,\mathcal{P}(k)
\simeq \frac{k_{\max}^2}{4\pi^2},
\end{equation}
and therefore
\begin{equation}
\langle |\chi|^2\rangle
=\frac{1}{2}\bigl(\langle \chi_1^2\rangle+\langle \chi_2^2\rangle\bigr)
\simeq \frac{k_{\max}^2}{4\pi^2}.
\end{equation}

In the simulations, the ultraviolet cutoff is fixed to \(k_{\max}=\eta\).

\subsection*{I.C Relation between \(M_\phi^2\), \(\mu_\phi^2\), and the vacuum energy}

The reduced effective potential for the real scalar field \(\phi\) is written as
\begin{equation}
U_{\rm re}(\phi)
=\frac{\lambda}{4}\phi^4-\frac{\beta}{3}\phi^3
+\frac{1}{2}M_\phi^2\phi^2,
\end{equation}
where the effective mass parameter \(M_\phi^2\) includes the contribution from fluctuations of the \(\chi\) field,
\begin{equation}
M_\phi^2=\mu_\phi^2+2g^2\langle|\chi|^2\rangle.
\end{equation}
Using the variance of the vacuum fluctuations given above and fixing
\(M_\phi^2=0.3\,\omega_*^2\),
we obtain
\begin{equation}
\mu_\phi^2 = 0.046697\,\omega_*^2 .
\end{equation}

The constant term \(U_0\) in the scalar potential is then fixed such that the energy density of the true vacuum vanishes. For the chosen parameter values, this yields
\begin{equation}
U_0 = 0.19965\,\omega_*^2\eta^2.
\end{equation}
This choice allows the mass of Q balls to be consistently determined from their total energy.

\subsection*{I.D Injection of a net \(U(1)\) charge}

The spatially averaged \(U(1)\) charge density is given by
\begin{equation}
\rho_{U(1)} \equiv \langle j^0\rangle
=2\,\langle \mathrm{Im}(\chi\,\pi_\chi)\rangle,
\end{equation}
where \(\pi_\chi\) denotes the canonical momentum conjugate to \(\chi\).
To obtain a nonvanishing net charge at the initial time, we initialize \(\pi_\chi\) as
\begin{equation}
\pi_\chi(\mathbf{x})
=\pi_{\rm stoc}(\mathbf{x})+\pi_\perp(\mathbf{x}),
\end{equation}
where \(\pi_{\rm stoc}\) is a stochastic component matched to the vacuum fluctuations of \(\chi\), and \(\pi_\perp\) is an additional component chosen to be locally orthogonal to \(\chi^*\) in field space.

Specifically, we take
\begin{equation}
\pi_\perp(\mathbf{x})
\equiv \frac{u_\perp}{\sqrt{2}}
\bigl[\chi_2(\mathbf{x})+i\,\chi_1(\mathbf{x})\bigr].
\end{equation}
With this choice, the stochastic component \(\pi_{\rm stoc}\) is statistically independent of \(\chi\), so that
\(\langle \chi\,\pi_{\rm stoc}\rangle=0\).
The spatially averaged charge density can then be estimated as
\begin{equation}
\rho_{U(1)}
\simeq u_\perp \langle |\chi|^2\rangle
\simeq u_\perp \frac{k_{\max}^2}{4\pi^2}.
\end{equation}
In this work we fix \(u_\perp = 0.1\,\omega_*\).

\subsection*{II. Stability criterion for a single Q-ball}

To quantitatively assess the stability of a single Q-ball formed during the evolution, we analyze the time dependence of its effective radius $R(t)$.
The effective radius is defined in the main text through the volume of the connected false-vacuum region.

To characterize both short-time fluctuations and possible long-term drift, we apply a sliding-window analysis to $R(t)$, where the radius is sampled at regular time intervals $\delta t = 10\,\omega_*^{-1}$.

For a time window of width $W$, we define the window-averaged radius
\begin{equation}
\mu_R=\frac{1}{W}\sum_{i=1}^{W} R_i ,
\end{equation}
and the corresponding standard deviation
\begin{equation}
\sigma_R=\sqrt{\frac{1}{W-1}\sum_{i=1}^{W} (R_i-\mu_R)^2 } .
\end{equation}
The relative fluctuation within each window is quantified by the coefficient of variation,
\begin{equation}
\mathrm{CV}_R=\frac{\sigma_R}{\mu_R} .
\end{equation}

To further detect possible secular evolution, we compare the mean radii in adjacent windows and define the relative drift as
\begin{equation}
d_R=\left|\frac{\mu_R^{\rm new}-\mu_R^{\rm old}}{\mu_R^{\rm old}}\right| .
\end{equation}

A configuration is identified as stable if both $\mathrm{CV}_R<\varepsilon_{\rm cv}$ and $d_R<\varepsilon_{\rm drift}$ are satisfied for several consecutive windows.
In the simulations presented in this work, we take $W=6$ and adopt the thresholds
$\varepsilon_{\rm cv}=5\times10^{-3}$ and $\varepsilon_{\rm drift}=2\times10^{-3}$.
We have verified that moderate variations of these parameters do not qualitatively affect the identification of stable Q-ball configurations.

\subsection*{III. Analytic estimate of the Q-ball mass density and numerical correction}

In this section we derive the analytic estimate for the Q-ball mass density and
define the correction factor extracted from numerical simulations.

We approximate the Q-ball number density by the density of isolated false-vacuum regions formed during the phase transition,
\begin{equation}
n_{\rm ball} \simeq \frac{N_b}{V}.
\end{equation}
The typical volume associated with a single Q-ball is therefore
\begin{equation}
V_* \equiv \frac{V}{N_b}.
\end{equation}

Assuming efficient charge collection (negligible leakage), the typical $U(1)$ charge per Q-ball is
\begin{equation}
Q_* = \Delta_\chi s(T_c)\,V_*,
\end{equation}
where $\Delta_\chi$ denotes the injected global $U(1)$ charge asymmetry normalized to the entropy density, and $s(T_c)$ is the entropy density at the percolation temperature $T_c$, such that the net $U(1)$ charge density is given by $\rho_{U(1)}=\Delta_\chi\, s(T_c)$.

For thin-wall Q-balls, the mass--charge relation yields
\begin{equation}
M_* = \frac{4\sqrt{2\pi}}{3}\,Q_*^{3/4}U_0^{1/4}.
\end{equation}
Combining the above expressions, the analytic estimate of the Q-ball mass density is
\begin{equation}
M_*\,n_{\rm ball}
=
\frac{4\sqrt{2\pi}}{3}\,
\bigl[\Delta_\chi s(T_c)\bigr]^{3/4}
\left(\frac{N_b\,U_0}{V}\right)^{1/4}.
\end{equation}

To compare with the numerical results, we define the simulation-based mass density
$\rho_{\mathrm{ball}}^{(\mathrm{num})}$.
Their ratio defines the correction factor
\begin{equation}
r \equiv \frac{\rho_{\mathrm{ball}}^{(\mathrm{num})}}{M_*\,n_{\rm ball}},
\end{equation}
which encapsulates the effects of the extended mass spectrum and partial charge leakage.

Using the relation $n_{\rm ball}(T_c)/s(T_c)=45/(2\pi^2 g_* T_c^3 V_*)$, the corrected Q-ball mass density normalized to the entropy density can be written as
\begin{equation}
\frac{\rho_{\mathrm{ball}}}{s}
=
r\,\frac{30\sqrt{2}}{\pi g_*}\,
\frac{U_0^{1/4}\,[\Delta_\chi s(T_c)]^{3/4}}{T_c^3\,V_*^{1/4}}.
\end{equation}

\end{document}